\documentclass[aps,pre,twocolumn,longbibliography,unsortedaddress,floatfix,10pt]{revtex4-1}

\usepackage{amsmath,amssymb,amsfonts}
\usepackage{algorithmic}
\usepackage{graphicx} 
\usepackage{color} 
\usepackage{textcomp}

\usepackage{mathtools}
\usepackage{bm}
\usepackage{url}

\newcommand{\BF}{\bm{F}}
\newcommand{\BH}{\bm{H}}
\newcommand{\BJ}{\bm{J}}
\newcommand{\BX}{\bm{X}}
\newcommand{\BZ}{\bm{Z}}
\newcommand{\Bdelta}{\boldsymbol{\delta}}
\newcommand{\calB}{\mathcal{B}}

\newcommand{\inner}[2]{#1 \cdot #2}
\newcommand{\average}[2]{\left\langle #1 \right\rangle_{#2}}

\newcommand{\norm}[1]{\left\| #1 \right\|}
\newcommand{\pd}[2]{\frac{\partial #1}{\partial #2}}

\begin{document}
\title{A Central Pattern Generator Network for Simple Control of Gait Transitions in Hexapod Robots based on Phase Reduction}

\author{Norihisa Namura}
\thanks{Corresponding author. E-mail: namura.n.aa@m.titech.ac.jp}
\affiliation{Department of Systems and Control Engineering, Tokyo Institute of Technology, Tokyo 152-8552, Japan}

\author{Hiroya Nakao}
\affiliation{Department of Systems and Control Engineering, Tokyo Institute of Technology, Tokyo 152-8552, Japan}
\affiliation{Research Center for Autonomous Systems Materialogy, Institute of Innovative Research, Tokyo Institute of Technology,
Yokohama, 226-8501, Japan
}

\date{\today}


\begin{abstract}
We present a model of the central pattern generator (CPG) network that can control gait transitions in hexapod robots in a simple manner based on phase reduction.
The CPG network consists of six weakly coupled limit-cycle oscillators, whose synchronization dynamics can be described by six phase equations through phase reduction.
Focusing on the transitions between the hexapod gaits with specific symmetries, the six phase equations of the CPG network can further be reduced to two independent equations for the phase differences.
By choosing appropriate coupling functions for the network, we can achieve desired synchronization dynamics regardless of the detailed properties of the limit-cycle oscillators used for the CPG.
The effectiveness of our CPG network is demonstrated by numerical simulations of gait transitions between the wave, tetrapod, and tripod gaits, using the FitzHugh-Nagumo oscillator as the CPG unit. 
\end{abstract}

\maketitle


\section{Introduction}

Legged locomotion is a key ability for many terrestrial organisms.
For instance, humans use bipedal locomotion, cats and horses use quadruped locomotion, and insects use hexapodal locomotion.
To adapt to various terrains or to change the speed of locomotion,
gait transitions often occur in vertebrates (e.g., horses~\cite{Heglund1988speed}) and invertebrates (e.g., drosophilas~\cite{Strauss1990coordination} and stick insects~\cite{Graham1981coordinated}).

Some neurophysiological experiments, such as the one for decerebrated cats~\cite{Shik1966control}, suggest that gait patterns 
are generated in the neuronal network called the \textit{central pattern generator} (CPG) in the spinal cord~\cite{Grillner1975locomotion}.
In invertebrates, CPGs also exist as distributed neuronal circuits that control gait patterns~\cite{Marder2005invertebrate,Grillner2006biological}.
The CPG can produce self-sustained rhythms for generating basic gait patterns without external stimulation or sensory feedback,
and it can further be modulated by the signals from the brain or sensory organs 
to switch between various rhythmic patterns, i.e., gait patterns~\cite{Rybak2006modelling}.

Compared to wheeled robots, legged robots have the advantage of adapting to uneven terrains and physical environments more efficiently,
and many studies have been conducted on biped robots~\cite{Aoi2005locomotion}, quadruped robots~\cite{Owaki2017quadruped}, hexapod robots~\cite{Ambe2013stability,Minati2018versatile}, and salamander robots~\cite{Ijspeert2007swimming}.
In particular, hexapod robots are more stable than biped or quadruped robots and have been applied, e.g., to humanitarian demining~\cite{Hexapod_demine} and exploration of the surface of Mars~\cite{Hexapod_mars}.
Since gait transitions are also important for hexapod robots, 
methods for realizing typical insect gaits and transitions between them using hexapod robots have been studied~\cite{Inagaki2006wave,Ricardo2010hexapod,Chen2012smooth,Manoonpong2013neural,Yu2016gait,Bal2021neural,Zhu2022generic}.

For controlling the locomotion of legged robots, CPG-based control strategy is an efficient bio-inspired method 
because it provides self-oscillations and requires less computational costs~\cite{Bal2021neural}.
CPG networks are widely modeled as coupled limit-cycle oscillators.
In rhythmic systems of mutually coupled limit-cycle oscillators like the CPG networks, synchronization can occur, leading to the alignment of their rhythms with each other. 
Synchronization of weakly coupled limit-cycle oscillators can be analyzed by the phase reduction theory~\cite{Kuramoto1984chemical,Hoppensteadt1997weakly,Winfree2001geometry,Ermentrout2010mathematical,Nakao2016phase},
which can approximately reduce the nonlinear multidimensional dynamics of the oscillator to a single-variable phase equation characterized by the natural frequency and phase response property of the oscillator.
Phase reduction can also be useful for engineering limit-cycle oscillators, 
e.g., in optimizing entrainment~\cite{Harada2010optimal,Dasanayake2011optimal,Zlotnik2013optimal,Monga2019phase,Kato2021optimization,Takata2021fast} and mutual synchronization~\cite{Shirasaka2017optimizing,Watanabe2019optimization,Namura2024optimal}.
Moreover, it has been used for analyzing gait-generation models~\cite{Ghigliazza2004minmal} and gait transitions~\cite{Aminzare2018gait,Yeldesbay2018role} in hexapods.

In this study, we present a CPG network for controlling gait transitions in hexapod robots, which allows various types of smooth limit-cycle oscillators to be used as the CPG unit.
We use phase reduction to derive six phase equations from the mathematical model of the CPG network consisting of six limit-cycle oscillators with weak coupling.
Using the symmetries of the typical hexapod gaits, i.e., the wave, tetrapod, and tripod gaits, we further reduce the six phase equations to two independent phase-difference equations.
By choosing appropriate coupling functions for the network, we can achieve all typical gaits irrespective of the detailed properties of the limit-cycle oscillators.
Compared to existing studies, our CPG network provides a simpler and more general framework in which the CPG model and coupling functions can be optimized.
We demonstrate the effectiveness of our CPG network by numerical simulations of gait transitions between the wave, tetrapod, and tripod gaits, adopting the FitzHugh-Nagumo oscillator as the CPG unit. 

This paper is organized as follows.
We first present a CPG network and explain the symmetric typical hexapod gaits in Sec.~\ref{sec:model}.
We next describe the gait control methods in Sec.~\ref{sec:methods}.
We demonstrate the results of the gait transitions by the present CPG network in Sec.~\ref{sec:results},
and we give conclusions in Sec.~\ref{sec:conclusions}.
Appendix~\ref{sec:gait_variations} gives further examples of the symmetric gait patterns.


\section{CPG Network}

\label{sec:model}

\begin{figure}
\centering
\includegraphics[width=0.4\textwidth]{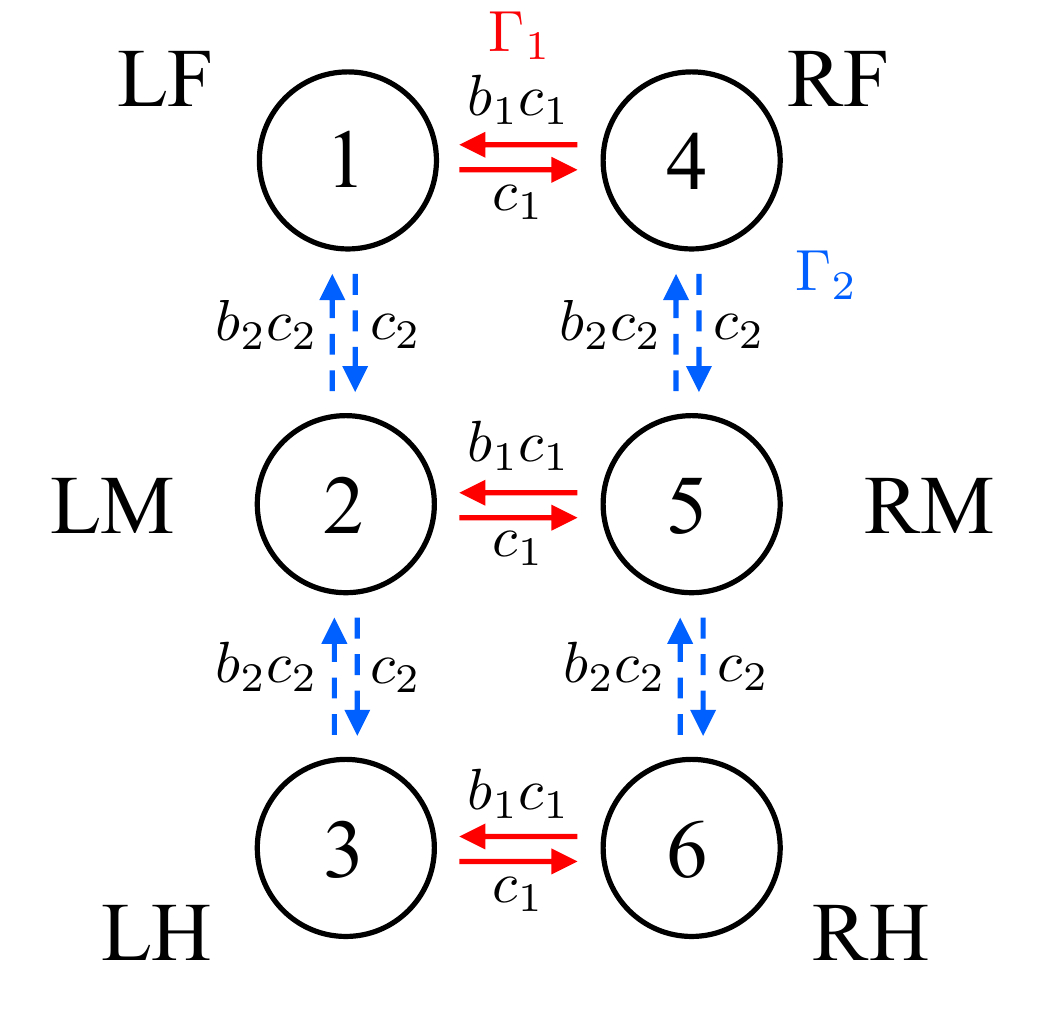}
\caption{
Coupling scheme of the present CPG network.
The six CPGs are denoted by 1 to 6, which correspond to LF, LM, LH, RF, RM, and RH, respectively.
The weights $c_{1,2}$ determine the coupling strength and $b_{1,2}$ determine whether the coupling is excitatory or inhibitory. 
The functions $\Gamma_{1,2}$ determine the phase coupling between the CPGs.
The red solid arrows show the coupling by $\Gamma_{1}$ and the blue dashed arrows show the coupling by $\Gamma_{2}$, respectively.
}
\label{fig1}
\end{figure}

In this section, we present a ladder-type bidirectional CPG network consisting of six limit-cycle oscillators as schematically shown in Fig.~\ref{fig1}.

\subsection{Phase dynamics of the CPG network}

We assume that the dynamics of the six limit-cycle oscillators in the CPG network (Fig.~\ref{fig1}) are described by
\begin{align}
\label{eq:model}
\begin{aligned}
\dot{\BX}_{1} ={}& s\BF(\BX_{1}) + \varepsilon \left( b_{1}c_{1}\tilde{\BH}_{1}(\BX_{1},\BX_{4}) + b_{2}c_{2}\tilde{\BH}_{2}(\BX_{1},\BX_{2}) \right), \\
\dot{\BX}_{2} ={}& s\BF(\BX_{2}) + \varepsilon b_{1}c_{1}\tilde{\BH}_{1}(\BX_{2},\BX_{5}) \\
&+ \varepsilon \left( c_{2}\tilde{\BH}_{2}(\BX_{2},\BX_{1}) + b_{2}c_{2}\tilde{\BH}_{2}(\BX_{2},\BX_{3}) \right), \\
\dot{\BX}_{3} ={}& s\BF(\BX_{3}) + \varepsilon \left( b_{1}c_{1}\tilde{\BH}_{1}(\BX_{3},\BX_{6}) + c_{2}\tilde{\BH}_{2}(\BX_{3},\BX_{2}) \right), \\
\dot{\BX}_{4} ={}& s\BF(\BX_{4}) + \varepsilon \left( c_{1}\tilde{\BH}_{1}(\BX_{4},\BX_{1}) + b_{2}c_{2}\tilde{\BH}_{2}(\BX_{4},\BX_{5}) \right), \\
\dot{\BX}_{5} ={}& s\BF(\BX_{5}) + \varepsilon c_{1}\tilde{\BH}_{1}(\BX_{5},\BX_{2}) \\
&+ \varepsilon \left( c_{2}\tilde{\BH}_{2}(\BX_{5},\BX_{4}) + b_{2}c_{2}\tilde{\BH}_{2}(\BX_{5},\BX_{6}) \right), \\
\dot{\BX}_{6} ={}& s\BF(\BX_{6}) + \varepsilon \left( c_{1}\tilde{\BH}_{1}(\BX_{6},\BX_{3}) + c_{2}\tilde{\BH}_{2}(\BX_{6},\BX_{5}) \right),
\end{aligned}
\end{align}
where $\BX_{k}(t) \in \mathbb{R}^{N}\; (k = 1,2,3,4,5,6)$ denotes each oscillator state at time $t$,
$\dot{\BX}$ represents the time derivative of $\BX$,
$\BF: \mathbb{R}^{N} \to \mathbb{R}^{N}$ is a smooth vector field common to all oscillators,
$s > 0$ is a common timescale parameter that determines the oscillation frequency of all the oscillators,
$c_{1,2} > 0$ are the coupling strength of $O(1)$, 
$b_{1,2} \in \{ -1, +1 \}$ determine whether the coupling is excitatory or inhibitory, 
$0 < \varepsilon \ll 1$ is a small parameter representing weakness of the coupling,
and $\tilde{\BH}_{1,2}: \mathbb{R}^{N} \times \mathbb{R}^{N} \to \mathbb{R}^{N}$ represent smooth coupling functions between the oscillators.
We assume that the original system $\dot\BX = \BF(\BX)$ without scaling ($s = 1$) has an exponentially stable limit cycle $\chi$ of natural period $T$ and frequency $\omega = 2\pi/T$,
where the state on $\chi$ at time $t$ is represented as $\tilde{\BX}_{0}(t)$ and satisfies $\tilde{\BX}_{0}(t) = \tilde{\BX}_{0}(t + T)$,
given the origin state $\tilde{\BX}_{0}(0)$ on the limit cycle.
Then, the period and frequency of the limit cycle $\chi$ of the time-rescaled vector field $s\BF$ are $T/s$ and $s\omega$, respectively.

The mathematical model of weakly coupled identical limit-cycle oscillators can be analyzed by phase reduction theory~\cite{Kuramoto1984chemical,Hoppensteadt1997weakly,Winfree2001geometry,Ermentrout2010mathematical,Nakao2016phase}.
We first introduce an asymptotic phase function $\Theta: \calB \to [0,2\pi)$ into the basin $\calB \subset \mathbb{R}^{N}$ of $\chi$
satisfying $\inner{\nabla\Theta(\BX)}{\BF(\BX)} = \omega$ for all $\BX \in \calB$,
where $\inner{\bm{a}}{\bm{b}} = \sum_{n=1}^{N} a_{n} b_{n}$ represents the scalar product of two vectors $\bm{a}$, $\bm{b} \in \mathbb{R}^{N}$.
Using the phase function, we can define the phase value $\theta$ of the oscillator state $\BX \in \calB$ by $\theta = \Theta(\BX)$, 
which increases constantly at the natural frequency $\omega$ over time,
\begin{align}
\dot{\theta}(t) = \inner{\nabla\Theta(\BX(t))}{\BF(\BX(t))} = \omega,
\end{align}
where we regard that the phase values $0$ and $2\pi$ are identical.
The state on $\chi$ can be represented by $\BX_{0}(\theta) = \tilde{\BX}_{0}(t = \theta/\omega)$ as a $2\pi$-periodic function of $\theta$, 
which satisfies $\BX_{0}(\theta) = \BX_{0}(\theta + 2\pi)$.

Since all coupling terms are of $O(\varepsilon)$ and sufficiently weak, we can assume that the state of each limit-cycle oscillator is close to $\chi$, i.e.,
\begin{align}
\label{eq:approx}
\BX_{k}(t) = \BX_{0}(\theta_{k}(t)) + O(\varepsilon) \quad (k = 1,2,3,4,5,6),
\end{align}
where $\theta_{k} = \Theta(\BX_{k})$ for $k = 1,2,3,4,5,6$.
Substituting Eq.~\eqref{eq:approx} into Eq.~\eqref{eq:model} and neglecting errors of $O(\varepsilon^{2})$, we can derive the following phase equations:
\begin{align}
\begin{aligned}
\dot{\theta}_{1} ={}& s\omega + \varepsilon \inner{\BZ(\theta_{1})}{\left( b_{1}c_{1}\BH_{1}(\theta_{1},\theta_{4}) + b_{2}c_{2}\BH_{2}(\theta_{1},\theta_{2}) \right)}, \\
\dot{\theta}_{2} ={}& s\omega + \varepsilon \inner{\BZ(\theta_{2})}{b_{1}c_{1}\BH_{1}(\theta_{2},\theta_{5})}, \\
&+ \varepsilon \inner{\BZ(\theta_{2})}{\left( c_{2}\BH_{2}(\theta_{2},\theta_{1}) + b_{2}c_{2}\BH_{2}(\theta_{2},\theta_{3}) \right)}, \\
\dot{\theta}_{3} ={}& s\omega + \varepsilon \inner{\BZ(\theta_{3})}{\left( b_{1}c_{1}\BH_{1}(\theta_{3},\theta_{6}) + c_{2}\BH_{2}(\theta_{3},\theta_{2}) \right)}, \\
\dot{\theta}_{4} ={}& s\omega + \varepsilon \inner{\BZ(\theta_{4})}{\left( c_{1}\BH_{1}(\theta_{4},\theta_{1}) + b_{2}c_{2}\BH_{2}(\theta_{4},\theta_{5}) \right)}, \\
\dot{\theta}_{5} ={}& s\omega + \varepsilon \inner{\BZ(\theta_{5})}{c_{1}\BH_{1}(\theta_{5},\theta_{2})}, \\
&+ \varepsilon \inner{\BZ(\theta_{5})}{\left( c_{2}\BH_{2}(\theta_{5},\theta_{4}) + b_{2}c_{2}\BH_{2}(\theta_{5},\theta_{6}) \right)}, \\
\dot{\theta}_{6} ={}& s\omega + \varepsilon \inner{\BZ(\theta_{6})}{\left( c_{1}\BH_{1}(\theta_{6},\theta_{3}) + c_{2}\BH_{2}(\theta_{6},\theta_{5}) \right)}.
\end{aligned}
\end{align}
Note that the frequency of each oscillator is scaled to $s \omega$.
Here, $\BZ: [0,2\pi) \to \mathbb{R}^{N}$ is the phase sensitivity function (PSF) of the oscillator defined by $\BZ(\theta) = \left. \nabla\Theta(\BX) \right|_{\BX = \BX_{0}(\theta)}$, which characterizes the linear response of the oscillator phase $\theta$ to small perturbations.
The PSF can be calculated as the $2\pi$-periodic solution to the following adjoint equation~\cite{Ermentrout2010mathematical,Brown2004phase}:
\begin{align}
\omega \frac{d}{d\theta}\BZ(\theta) = -\BJ(\BX_{0}(\theta))^{\top}\BZ(\theta),
\end{align}
where $\BJ: \mathbb{R}^{N} \to \mathbb{R}^{N \times N}$ is the Jacobian matrix of the vector field $\BF$, i.e., $\BJ(\BX) = \nabla \BF(\BX)$.
The PSF satisfies the following normalization condition:
\begin{align}
\inner{\BZ(\theta)}{\frac{d\BX_{0}(\theta)}{d\theta}} = 1.
\end{align}
The functions $\BH_{1,2}: [0,2\pi) \times [0,2\pi) \to \mathbb{R}^{N}$ are mutual coupling functions (MCFs) defined by
\begin{align}
\BH_{1,2}(\theta_{i},\theta_{j}) &= \tilde{\BH}_{1,2}(\BX_{0}(\theta_{i}),\BX_{0}(\theta_{j})),
\end{align}
which depend on two phases $\theta_{i}$ and $\theta_{j}$ for $i,j \in \{ 1,2,3,4,5,6 \}$.
From Eq.~\eqref{eq:approx}, 
\begin{align}
\tilde{\BH}_{1,2}(\BX_{i},\BX_{j}) &= \BH_{1,2}(\theta_{i},\theta_{j}) + O(\varepsilon) 
\end{align}
holds for $i,j \in \{ 1,2,3,4,5,6 \}$.

Since the frequency of the oscillator is $s \omega$ and therefore the relative phase $\theta_{k} - s\omega t$ is a slow variable, 
we can conduct the averaging approximation~\cite{Kuramoto1984chemical,Hoppensteadt1997weakly,Nakao2016phase}
over one period $T/s$ and obtain the following averaged phase equations by neglecting errors of $O(\varepsilon^{2})$:
\begin{align}
\label{eq:phase_eq}
\begin{aligned}
\dot{\theta}_{1} ={}& s\omega + \varepsilon (b_{1}c_{1}\Gamma_{1}(\theta_{1} - \theta_{4}) + b_{2}c_{2}\Gamma_{2}(\theta_{1} - \theta_{2})), \\
\dot{\theta}_{2} ={}& s\omega + \varepsilon b_{1}c_{1}\Gamma_{1}(\theta_{2} - \theta_{5}) \\
&+ \varepsilon (c_{2}\Gamma_{2}(\theta_{2} - \theta_{1}) + b_{2}c_{2}\Gamma_{2}(\theta_{2} - \theta_{3})), \\
\dot{\theta}_{3} ={}& s\omega + \varepsilon (b_{1}c_{1}\Gamma_{1}(\theta_{3} - \theta_{6}) + c_{2}\Gamma_{2}(\theta_{3} - \theta_{2})), \\
\dot{\theta}_{4} ={}& s\omega + \varepsilon (c_{1}\Gamma_{1}(\theta_{4} - \theta_{1}) + b_{2}c_{2}\Gamma_{2}(\theta_{4} - \theta_{5})), \\
\dot{\theta}_{5} ={}& s\omega + \varepsilon c_{1}\Gamma_{1}(\theta_{5} - \theta_{2}) \\
&+ \varepsilon (c_{2}\Gamma_{2}(\theta_{5} - \theta_{4}) + b_{2}c_{2}\Gamma_{2}(\theta_{5} - \theta_{6})), \\
\dot{\theta}_{6} ={}& s\omega + \varepsilon (c_{1}\Gamma_{1}(\theta_{6} - \theta_{3}) + c_{2}\Gamma_{2}(\theta_{6} - \theta_{5})), \\
\end{aligned}
\end{align}
where the functions $\Gamma_{1,2}$ are called phase coupling functions (PCFs), which are $2\pi$-periodic functions represented by 
\begin{align}
\Gamma_{1,2}(\varphi) = \average{\inner{\BZ(\psi)}{\BH_{1,2}(\psi,\psi - \varphi)}}{\psi}.
\end{align}
Here, we denote the averaging of a smooth $2\pi$-periodic function $g$ over one period of oscillation as 
\begin{align}
\average{g(\psi)}{\psi} = \frac{1}{2\pi} \int_{0}^{2\pi} g(\psi) d\psi.
\end{align}

Now, we assume that the MCFs are given in the following specific form: 
\begin{align}
\label{eq:H12}
\BH_{1,2}(\theta_{i},\theta_{j}) = \frac{\BZ(\theta_{i}) P_{1,2}(\theta_{i} - \theta_{j})}{\average{\norm{\BZ(\psi)}^{2}}{\psi}},
\end{align}
where $P_{1,2}: \mathbb{R} \to \mathbb{R}$ are arbitrary $2\pi$-periodic smooth functions.
As discussed in~\cite{Namura2024optimal}, these MCFs give the most energy-efficient forms of the mutual coupling that maximize the absolute value of the corresponding PCFs at each value of the phase difference under a given power, leading to fast mutual synchronization of the oscillators.
Introducing $\varphi = \theta_{i} - \theta_{j}$, the PCFs are obtained from the above MCFs as
\begin{align}
\label{eq:Gamma_P}
\Gamma_{1,2}(\varphi) = P_{1,2}(\varphi).
\end{align}
By choosing the functional forms of $P_{1,2}(\varphi)$, we can realize any forms of $\Gamma_{1,2}$.
Thus, we need to consider only the PCFs $\Gamma_{1,2}$ in designing the CPG network, and then MCFs $\BH_{1,2}$ follow from Eq.~\eqref{eq:H12}.
We note that, in general, the functional form of the PCF depends on both the PSF and MCF.
For the above particular MCF, the PCF become independent of the PSF and we can arbitrarily design it to realize desired synchronization dynamics.


\subsection{Symmetric gait patterns}

Next, we describe the hexapod gaits considered in this study.
The six legs of the hexapod robots are called the left front (LF), left middle (LM), left hind (LH), right front (RF), right middle (RM), and right hind (RH), as shown in Fig.~\ref{fig1}.
We consider $(\theta_{1},\theta_{2},\theta_{3},\theta_{4},\theta_{5},\theta_{6})$ to be the phases of (LF, LM, LH, RF, RM, RH), respectively.

We assume that the hexapod gaits have the following symmetries:
\begin{align}
\label{eq:contra_sym}
\theta_{6} - \theta_{3} = \theta_{5} - \theta_{2} = \theta_{4} - \theta_{1} &\mod 2\pi, \\
\label{eq:ipsi_sym}
\theta_{3} - \theta_{2} = \theta_{2} - \theta_{1} = \theta_{6} - \theta_{5} = \theta_{5} - \theta_{4} &\mod 2\pi, 
\begin{aligned}
\end{aligned}
\end{align}
where Eq.~\eqref{eq:contra_sym} indicates that the phase differences between the contralateral legs in the front, middle, and hind positions are equal, 
and Eq.~\eqref{eq:ipsi_sym} indicates that the phase differences between the adjacent ipsilateral legs are equal, i.e.,
the phase shift of the middle legs from the front legs is the same as that of the hind legs from the middle legs on both sides.

\begin{figure}
\centering
\includegraphics[width=0.4\textwidth]{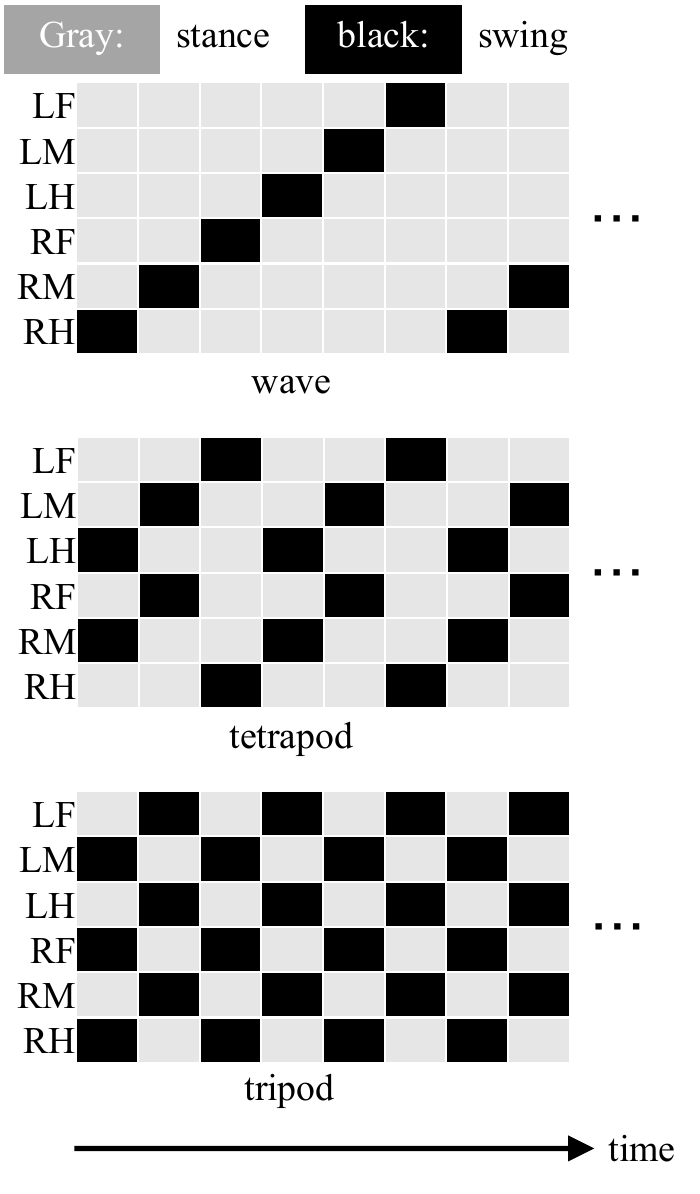}
\caption{Diagrams of the wave, tetrapod, and tripod gaits (see Sec.~\ref{sec:methods}~A for the definition of the swing and stance phases). 
}
\label{fig2}
\end{figure}

As an example, we consider three typical hexapod gaits, i.e., the slowest gait called the \textit{wave gait}, the middle-speed gait called the \textit{tetrapod gait}, and the fastest gait called the \textit{tripod gait}~\cite{Wilson1966insect,Collins1993hexapodal,Golubitsky1998modular}.
Figure~\ref{fig2} shows the diagrams of the three gaits, where the states (stance or swing) of the six legs are plotted in gray and black colors, respectively.
In the wave gait (also called a metachronal gait), the hexapod robot moves the six legs, i.e., (RH), (RM), (RF), (LH), (LM), and (LF), in order at equal phase intervals of $\pi/3$.
In the tetrapod gait, the legs form three synchronized pairs, i.e., (LH, RM), (LM, RF), and (LF, RH), which are moved in order at equal phase intervals of $2\pi/3$.
Finally, in the tripod gait, the legs form two synchronized triplets, (RF, LM, RH) and (LF, RM, LH), which are alternatively moved (i.e., with the phase interval of $\pi$).

The phase shifts of the legs in the three types of the gaits correspond to
\begin{align}
\begin{aligned}
&(\theta_{1},\theta_{2},\theta_{3},\theta_{4},\theta_{5},\theta_{6}) \\
&= \left\{
\begin{array}{ll}
\left( 0, \frac{1}{3}\pi, \frac{2}{3}\pi, \pi,\frac{4}{3}\pi, \frac{5}{3}\pi \right) \quad (\textrm{wave}), \\\\
\left( 0, \frac{2}{3}\pi, \frac{4}{3}\pi, \frac{2}{3}\pi, \frac{4}{3}\pi, 0 \right) \quad (\textrm{tetrapod}), \\\\
\left( 0, \pi, 0, \pi, 0, \pi \right) \quad (\textrm{tripod}),
\end{array}
\right. 
\end{aligned}
\end{align}
when the LF is chosen as the reference, $\theta_{1} = 0$.
We can find that all the three gaits satisfy the symmetries given in Eqs.~\eqref{eq:contra_sym} and \eqref{eq:ipsi_sym}.
We note that other hexapod gaits are also possible as described in Appendix~\ref{sec:gait_variations}.

\subsection{Equations for the phase differences}

We now show that the six phase equations can be reduced to two independent phase-difference equations by using the symmetries~\eqref{eq:contra_sym} and \eqref{eq:ipsi_sym} of the gaits.
First, we consider the symmetry of the phase differences between the contralateral legs given in Eq.~\eqref{eq:contra_sym}, which yields
\begin{align}
\label{eq:phase3_diff}
\theta_{k+3} - \theta_{m+3} = \theta_{k} - \theta_{m} \mod 2\pi \quad (k,m = 1,2,3).
\end{align}
Plugging Eqs.~\eqref{eq:contra_sym} and \eqref{eq:phase3_diff} into Eq.~\eqref{eq:phase_eq}, we find that the time derivatives of $\theta_{k+3} - \theta_{k}$ take the same value for $k = 1,2,3$, i.e., 
\begin{align}
\label{eq:phase3_diff_dynamics}
\dot{\theta}_{k+3} - \dot{\theta}_{k} = \varepsilon c_{1} \left( \Gamma_{1}(\theta_{k+3} - \theta_{k}) - b_{1}\Gamma_{1}(\theta_{k} - \theta_{k+3}) \right).
\end{align}
Therefore, if the initial phases $\{ \theta_{k}(0) \}_{k=1}^{6}$ satisfy the condition~\eqref{eq:contra_sym}, the phases satisfy 
\begin{align}
\label{eq:left_right}
\theta_{k+3}(t) - \theta_{k}(t) \coloneqq \alpha(t) \mod 2\pi \quad (k = 1,2,3)
\end{align}
for all $t$, i.e., the phase differences 
$\theta_{4} - \theta_{1}$, $\theta_{5} - \theta_{2}$, and $\theta_{6} - \theta_{3}$ are always equal to $\alpha$.
Therefore, the time evolution of the phase difference $\alpha$ obeys
\begin{align}
\dot{\alpha} = \varepsilon c_{1} \left( \Gamma_{1}(\alpha) - b_{1}\Gamma_{1}(-\alpha) \right).
\end{align}

From Eqs.~\eqref{eq:phase3_diff} and \eqref{eq:left_right}, by using the phase difference $\alpha$, the dynamics of the six phases can be rewritten as
\begin{align}
\begin{aligned}
\dot{\theta}_{1} &= s\omega + \varepsilon (b_{1}c_{1}\Gamma_{1}(-\alpha) + b_{2}c_{2}\Gamma_{2}(\theta_{1} - \theta_{2})), \\
\dot{\theta}_{2} &= s\omega + \varepsilon (b_{1}c_{1}\Gamma_{1}(-\alpha) + c_{2}\Gamma_{2}(\theta_{2} - \theta_{1}) + b_{2}c_{2}\Gamma_{2}(\theta_{2} - \theta_{3})), \\
\dot{\theta}_{3} &= s\omega + \varepsilon (b_{1}c_{1}\Gamma_{1}(-\alpha) + c_{2}\Gamma_{2}(\theta_{3} - \theta_{2})), \\
\dot{\theta}_{4} &= s\omega + \varepsilon (c_{1}\Gamma_{1}(\alpha) + b_{2}c_{2}\Gamma_{2}(\theta_{1} - \theta_{2})), \\
\dot{\theta}_{5} &= s\omega + \varepsilon (c_{1}\Gamma_{1}(\alpha) + c_{2}\Gamma_{2}(\theta_{2} - \theta_{1}) + b_{2}c_{2}\Gamma_{2}(\theta_{2} - \theta_{3})), \\
\dot{\theta}_{6} &= s\omega + \varepsilon (c_{1}\Gamma_{1}(\alpha) + c_{2}\Gamma_{2}(\theta_{3} - \theta_{2})).
\end{aligned}
\end{align}

We next consider the symmetry of the phase differences between the ipsilateral legs, given in Eq.~\eqref{eq:ipsi_sym}.
If the initial phases $\{ \theta_{k}(0) \}_{k=1}^{6}$ satisfy the symmetry~\eqref{eq:ipsi_sym} and it is maintained during the subsequent evolution,
their phase differences 
$\varphi_{12} = \theta_{1} - \theta_{2}$, $\varphi_{32} = \theta_{3} - \theta_{2}$, $\varphi_{45} = \theta_{4} - \theta_{5}$, and $\varphi_{65} = \theta_{6} - \theta_{5}$
satisfy
\begin{align}
\label{eq:beta}
-\varphi_{12}(t) = \varphi_{32}(t) = -\varphi_{45}(t) = \varphi_{65}(t) \coloneqq \beta(t) \mod 2\pi
\end{align}
for all $t$.
We only need to consider the dynamics of two phase differences $\varphi_{12} = \theta_{1} - \theta_{2}$ and $\varphi_{32} = \theta_{3} - \theta_{2}$ because 
$\varphi_{65} = \varphi_{32}$ and $\varphi_{54} = \varphi_{21}$ from Eq.~\eqref{eq:phase3_diff}, 
which are given by
\begin{align}
\label{eq:varphi12}
\dot{\varphi}_{12} &= \varepsilon \left( b_{2}c_{2}\Gamma_{2}(\varphi_{12}) - c_{2}\Gamma_{2}(-\varphi_{12}) - b_{2}c_{2}\Gamma_{2}(-\varphi_{32}) \right), \\
\label{eq:varphi32}
\dot{\varphi}_{32} &= \varepsilon \left( c_{2}\Gamma_{2}(\varphi_{32}) - c_{2}\Gamma_{2}(-\varphi_{12}) - b_{2}c_{2}\Gamma_{2}(-\varphi_{32}) \right),
\end{align}
respectively. 
Since $-\varphi_{12} = \varphi_{32} = \beta$, the dynamics of the phase difference $\beta$ can be represented as
\begin{align}
\label{eq:beta_relationship}
\dot{\beta} = \varepsilon c_{2}\Gamma_{2}(\beta) = -\varepsilon b_{2}c_{2}\Gamma_{2}(-\beta) 
\end{align}
from Eqs.~\eqref{eq:varphi12} and \eqref{eq:varphi32}.

For $-\varphi_{12}$ and $\varphi_{32}$ to remain equal, Eq.~\eqref{eq:beta_relationship} should hold for all $\beta$,
where the PCF $\Gamma_{2}$ should be an odd or even function because $\Gamma_{2}$ is a continuous nonzero $2\pi$-periodic function and $c_{2}$ is a non-zero constant. 
The parameter $b_{2}$ determining excitatory or inhibitory coupling should be chosen as 
\begin{align}
\label{cond:b2}
\left\{
\begin{array}{ll}
b_{2} = 1 & \quad \mathrm{if\;} \Gamma_{2} \mathrm{\; is\; an\; odd\; function}, \\ 
b_{2} = -1 & \quad \mathrm{if\;} \Gamma_{2} \mathrm{\; is\; an\; even\; function}.
\end{array}
\right. 
\end{align}

Thus, the dynamics of the six phases can be described by two independent equations for the phase differences $\alpha$ between the contralateral legs and $\beta$ between the adjacent ipsilateral legs,
\begin{align}
\label{eq:Gamma_alpha}
\dot{\alpha} &= \varepsilon c_{1} ( \Gamma_{1}(\alpha) - b_{1}\Gamma_{1}(-\alpha) ) \coloneqq \varepsilon \Gamma_{\mathrm{alpha}}(\alpha), \\
\label{eq:Gamma_beta}
\dot{\beta} &= \varepsilon c_{2}\Gamma_{2}(\beta) \coloneqq \varepsilon \Gamma_{\mathrm{beta}}(\beta).
\end{align}
By controlling $\alpha$ and $\beta$, we can realize different gait patterns.
The fixed point $(\alpha^{*},\beta^{*})$ of Eqs.~\eqref{eq:Gamma_alpha} and \eqref{eq:Gamma_beta} for each gait is 
\begin{align}
(\alpha^{*},\beta^{*}) = \left\{
\begin{array}{ll}
\left( \pi, \frac{1}{3}\pi \right) \quad (\textrm{wave}), \\\\
\left( \frac{2}{3}\pi, \frac{2}{3}\pi \right) \quad (\textrm{tetrapod}), \\\\
\left( \pi, \pi \right) \quad (\textrm{tripod}).
\end{array}
\right. 
\end{align}
For these fixed points corresponding to the individual gaits to be realizable, they should be linearly stable.
Moreover, we need the stability of the periodic solution satisfying the symmetries~\eqref{eq:contra_sym} and \eqref{eq:ipsi_sym}.
These will be analyzed in the next section.
 

\subsection{Stability analysis}

We first discuss the stability conditions for the periodic solution of the CPG network dynamics satisfying the symmetries~\eqref{eq:contra_sym} and \eqref{eq:ipsi_sym}.
We then describe the conditions for the fixed point $(\alpha^{*}, \beta^{*})$ to ensure that the gaits are stable.

For the periodic solution of the CPG to maintain the symmetry~\eqref{eq:contra_sym}, the variable $\Bdelta = \left[ \delta_{1}\; \delta_{2} \right]$ defined by 
\begin{align}
\delta_{1} &= \varphi_{41} - \varphi_{52}, \\
\delta_{2} &= \varphi_{63} - \varphi_{52}
\end{align}
should be stable at the fixed point $\Bdelta = [0\; 0] \mod 2\pi$, where $\varphi_{41} = \theta_{4} - \theta_{1}$, $\varphi_{52} = \theta_{5} - \theta_{2}$, and $\varphi_{63} = \theta_{6} - \theta_{3}$.
Hereafter in this section, for simplicity, we omit writing `mod $2\pi$' for the fixed points.
From Eq.~\eqref{eq:phase3_diff_dynamics}, the dynamics of $\delta_{1,2}$ are represented as
\begin{align}
\begin{aligned}
\dot{\delta}_{1} ={}& \dot{\varphi}_{41} - \dot{\varphi}_{52} \\
={}& c_{1}\Gamma_{1}(\varphi_{41}) - b_{1}c_{1}\Gamma_{1}(-\varphi_{41}) \\
&- c_{1}\Gamma_{1}(\varphi_{52}) + b_{1}c_{1}\Gamma_{1}(-\varphi_{52}), 
\end{aligned}
\end{align}
and
\begin{align}
\begin{aligned}
\dot{\delta}_{2} ={}& \dot{\varphi}_{63} - \dot{\varphi}_{52} \\
={}& c_{1}\Gamma_{1}(\varphi_{63}) - b_{1}c_{1}\Gamma_{1}(-\varphi_{63}) \\
&- c_{1}\Gamma_{1}(\varphi_{52}) + b_{1}c_{1}\Gamma_{1}(-\varphi_{52}), 
\end{aligned}
\end{align}
respectively.
The fixed point $\Bdelta = 0$ is linearly stable if all eigenvalues of 
the Jacobian $\partial \dot{\Bdelta} / \partial \Bdelta$ have negative real parts
at $\Bdelta = 0$.
By using the chain rule of derivatives, we obtain
\begin{align}
\begin{aligned}
\pd{\dot{\Bdelta}}{\Bdelta} &=
\begin{bmatrix}
\pd{\dot{\delta}_{1}}{\delta_{1}} & \pd{\dot{\delta}_{1}}{\delta_{2}} \\
\pd{\dot{\delta}_{2}}{\delta_{1}} & \pd{\dot{\delta}_{2}}{\delta_{2}} 
\end{bmatrix}
\\ &=
\begin{bmatrix}
\pd{\dot{\delta}_{1}}{\varphi_{41}} - \pd{\dot{\delta}_{1}}{\varphi_{52}} & \pd{\dot{\delta}_{1}}{\varphi_{63}} - \pd{\dot{\delta}_{1}}{\varphi_{52}} \\
\pd{\dot{\delta}_{2}}{\varphi_{41}} - \pd{\dot{\delta}_{2}}{\varphi_{52}} & \pd{\dot{\delta}_{2}}{\varphi_{63}} - \pd{\dot{\delta}_{2}}{\varphi_{52}}
\end{bmatrix}
\\ &= \varepsilon (c_{1}\Gamma'_{1}(\alpha) + b_{1}c_{1}\Gamma'_{1}(-\alpha))
\begin{bmatrix}
2 & 1 \\ 1 & 2
\end{bmatrix}
.
\end{aligned}
\end{align}
Therefore, the eigenvalues of $\partial \dot{\Bdelta} / \partial \Bdelta$ are negative if 
\begin{align}
\label{eq:stability_alpha_sym}
\varepsilon (c_{1}\Gamma'_{1}(\alpha) + b_{1}c_{1}\Gamma'_{1}(-\alpha) ) < 0.
\end{align}
If we use an odd or even function as $\Gamma_{1}$, this condition can be rewritten as 
\begin{align}
\varepsilon c_{1}\Gamma'_{1}(\alpha) < 0
\end{align}
by taking $b_{1}$ as
\begin{align}
\label{cond:b1}
\left\{
\begin{array}{ll}
b_{1} = 1 & \quad \mathrm{if\;} \Gamma_{1} \mathrm{\; is\; an\; odd\; function}, \\ 
b_{1} = -1 & \quad \mathrm{if\;} \Gamma_{1} \mathrm{\; is\; an\; even\; function}.
\end{array}
\right.
\end{align}
Since $\varepsilon > 0$ and $c_{1} > 0$, $\Bdelta = \bm{0}$ is stable if 
\begin{align}
\label{cond:alpha_sym_stable}
\Gamma'_{1}(\alpha) < 0.
\end{align}
This condition cannot be satisfied for the whole range of $\alpha \in [0,2\pi)$, but
it is sufficient if this condition holds in an interval $\alpha \in [\alpha_{\mathrm{init}}, \alpha^{*}]$ (or $[\alpha^{*}, \alpha_{\mathrm{init}}]$),
where $\alpha_{\mathrm{init}}$ and $\alpha^{*}$ are the initial value and the fixed point of $\alpha$, respectively.

From now on, we assume that the above condition is satisfied.
For the symmetry~\eqref{eq:ipsi_sym}, or equivalently~\eqref{eq:beta} of the periodic solution to be kept, the variable $\delta_{3}$ defined by
\begin{align}
\delta_{3} = \varphi_{12} + \varphi_{32}
\end{align}
should be stable at the fixed point $\delta_{3} = 0$.
By adding Eqs.~\eqref{eq:varphi12} and \eqref{eq:varphi32} and using Eq.~\eqref{eq:beta_relationship},
the rate of change $\dot{\delta}_{3}$ at $\delta_{3} = 0$ satisfies
\begin{align}
\label{eq:delta_dynamics}
\begin{aligned}
\dot{\delta}_{3} ={}& \varepsilon \left( b_{2}c_{2}\Gamma_{2}(\varphi_{12}) - 2c_{2}\Gamma_{2}(-\varphi_{12}) \right) \\
&+ \varepsilon \left( c_{2}\Gamma_{2}(\varphi_{32}) - 2b_{2}c_{2}\Gamma_{2}(-\varphi_{32}) \right) \\
={}& \varepsilon c_{2} ( -\Gamma_{2}(\beta) - b_{2}\Gamma_{2}(-\beta) ) \\
={}& 0.
\end{aligned}
\end{align}
The variable $\delta_{3} = 0$ is linearly stable if 
\begin{align}
\begin{aligned}
\pd{\dot{\delta}_{3}}{\delta_{3}} ={}& 
\pd{\dot{\delta}_{3}}{\varphi_{12}} + \pd{\dot{\delta}_{3}}{\varphi_{32}} \\
={}& \varepsilon \left( b_{2}c_{2}\Gamma'_{2}(\varphi_{12}) + 2c_{2}\Gamma'_{2}(-\varphi_{12}) \right) \\
&+ \varepsilon \left( c_{2}\Gamma'_{2}(\varphi_{32}) + 2b_{2}c_{2}\Gamma'_{2}(-\varphi_{32}) \right) \\
={}& 3\varepsilon c_{2} ( \Gamma'_{2}(\beta) + b_{2}\Gamma'_{2}(-\beta) ) < 0,
\end{aligned}
\end{align}
where we used the chain rule of derivatives.
This quantity can be calculated as
\begin{align}
\pd{\dot{\delta}_{3}}{\delta_{3}} = 6\varepsilon c_{2} \Gamma'_{2}(\beta) 
\end{align}
for both odd and even functions of $\Gamma_{2}$ using the condition~\eqref{cond:b2}.
Since $\varepsilon > 0$ and $c_{2} > 0$, $\delta_{3} = 0$ is stable if 
\begin{align}
\label{cond:beta_sym_stable}
\Gamma'_{2}(\beta) < 0.
\end{align}
Although this condition does not hold for all $\beta \in [0,2\pi)$, 
it is sufficient if this condition holds in the interval $\beta \in [\beta_{\mathrm{init}}, \beta^{*}]$ (or $[\beta^{*}, \beta_{\mathrm{init}}]$),
where $\beta_{\mathrm{init}}$ and $\beta^{*}$ are the initial value and the fixed point, respectively.

We have shown the conditions for the linear stability of the symmetric periodic orbit. 
We now focus on the stability of the fixed point $(\alpha^{*}, \beta^{*})$. 
The fixed point $\alpha = \alpha^{*}$ is linearly stable if the following conditions hold:
\begin{align}
\Gamma_{\mathrm{alpha}}(\alpha^{*}) &= c_{1} (\Gamma_{1}(\alpha^{*}) - b_{1}\Gamma_{1}(-\alpha^{*})) = 0, \\
\label{eq:alpha_stability}
\Gamma'_{\mathrm{alpha}}(\alpha^{*}) &= c_{1} (\Gamma'_{1}(\alpha^{*}) + b_{1}\Gamma'_{1}(-\alpha^{*})) < 0.
\end{align}
If we use an odd or even function for $\Gamma_{1}$ and choose $b_{1}$ as given by the condition~\eqref{cond:b1},
\begin{align}
\Gamma_{1}(\alpha^{*}) &= 0, \\
\label{cond:stability1}
\Gamma'_{1}(\alpha^{*}) &< 0
\end{align}
are the conditions for the linear stability.
The stability condition~\eqref{cond:stability1} for the fixed point $\alpha^{*}$ is consistent with the condition~\eqref{cond:alpha_sym_stable} for the symmetry~\eqref{eq:contra_sym}.

The fixed point $\beta = \beta^{*}$ is linearly stable if the following conditions hold:
\begin{align}
\Gamma_{2}(\beta^{*}) &= 0, \\
\label{cond:stability2}
\Gamma'_{2}(\beta^{*}) &< 0,
\end{align}
which follow from $c_{2} > 0$ and Eq.~\eqref{eq:Gamma_beta}.
The stability condition~\eqref{cond:stability2} for the fixed point $\beta^{*}$ is consistent with the condition~\eqref{cond:beta_sym_stable} for the symmetry~\eqref{eq:ipsi_sym}.


\section{Gait control methods}

\label{sec:methods}

In this section, we explain the methods for gait control.
Each gait is generated by moving the legs when the corresponding phase values exceed a threshold, where the timing is adjusted by a mutually synchronized solution of the CPGs with appropriate PCFs $\Gamma_{1,2}$.
Gait transitions can be conducted by changing the PCFs $\Gamma_{1,2}$ between odd and even functions to switch the fixed points and by alternating $b_{1,2}$ according to the conditions \eqref{cond:b1} and \eqref{cond:b2}. 
The timescale parameter $s$ is adjusted so that the duration of the swing phase is the same for all of the wave, tetrapod, and tripod gaits.


\subsection{Determining gaits from CPG outputs}

We first describe the method for determining the swing or stance phase from the CPG output.
We choose one component $x_{\mathrm{out}}$ of the CPG unit (oscillator) as the output, although the oscillator itself is multidimensional.
The corresponding leg is in the swing phase if the output $x_{\mathrm{out}}$ is above a threshold $\sigma$ and in the stance phase otherwise~\cite{Yu2016gait}, i.e.,
\begin{align}
\label{cond:swing_stance}
\left\{
\begin{array}{ll}
\mathrm{swing} \quad &\mathrm{if} \quad x_{\mathrm{out}} > \sigma, \\
\mathrm{stance} \quad &\mathrm{otherwise}.
\end{array}
\right. 
\end{align}
We assume the output $x_{\mathrm{out}}$ to have only two extremum points within one period. 
The threshold $\sigma$ for each gait is determined to realize the following duty factor $r_{\mathrm{DF}}$:
\begin{align}
r_{\mathrm{DF}} = \frac{T_{\mathrm{st}}}{T_{\mathrm{sw}} + T_{\mathrm{st}}},
\end{align}
where $T_{\mathrm{sw}}$ and $T_{\mathrm{st}}$ represent the durations of the swing phase and stance phase in one gait period $T/s$, respectively, 
where $T_{\mathrm{sw}} + T_{\mathrm{st}} = T/s$ holds.
The duty factors for the wave, tetrapod, and tripod gaits are chosen as $5/6$, $2/3$, and $1/2$, respectively.

The timescale parameter $s$ is chosen as
\begin{align}
\label{eq:speed}
s = \left\{
\begin{array}{ll}
\frac{1}{6} \quad (\textrm{wave}), \\\\
\frac{1}{3} \quad (\textrm{tetrapod}), \\\\
\frac{1}{2} \quad (\textrm{tripod}),
\end{array}
\right. 
\end{align}
so that the duration of the swing phase of each leg is the same for all of the wave, tetrapod, and tripod gaits.


\subsection{Functional forms of the coupling functions}

We here describe the odd function $\Gamma_{\mathrm{odd}}$ and even function $\Gamma_{\mathrm{even}}$ that we use for gait transitions.
Both of these functions are necessary, because for the tripod gait with the fixed point $(\alpha^{*}, \beta^{*}) = (\pi, \pi)$,
any $2\pi$-periodic odd function $\Gamma_{\mathrm{odd}}(\varphi)$ has a fixed point at $\varphi = \pi$, preventing the tripod gait from switching to other gaits,
and any $2\pi$-periodic even function $\Gamma_{\mathrm{even}}(\varphi)$ satisfies $\Gamma'_{\mathrm{even}}(\pi) = 0$, where the linear stability of $\varphi = \pi$ is not guaranteed, i.e., tripod gait is not linearly stable.
We thus switch the odd and even functions to realize gait transitions.

Since the fixed points of the three gaits are  $(\alpha^{*}, \beta^{*}) =$ $(\pi, \pi/3)$, $(2\pi/3, 2\pi/3)$, and $(\pi, \pi)$,
there can be six cases of transitions:
\begin{align}
\begin{matrix}
\frac{1}{3}\pi \to \frac{2}{3}\pi, & \frac{2}{3}\pi \to \frac{1}{3}\pi, \\\\
\frac{1}{3}\pi \to \pi, & \pi \to \frac{1}{3}\pi, \\\\
\frac{2}{3}\pi \to \pi, & \pi \to \frac{2}{3}\pi.
\end{matrix}
\end{align}
Here, we choose the odd function as the PCFs $\Gamma_{1,2}$ for the transitions $\pi/3 \to \pi$ and $2\pi/3 \to \pi$ because the $2\pi$-periodic odd function always has a fixed point at $\pi$. 
We choose the even function for other transitions.

\begin{figure}
\centering
\includegraphics[width=0.48\textwidth]{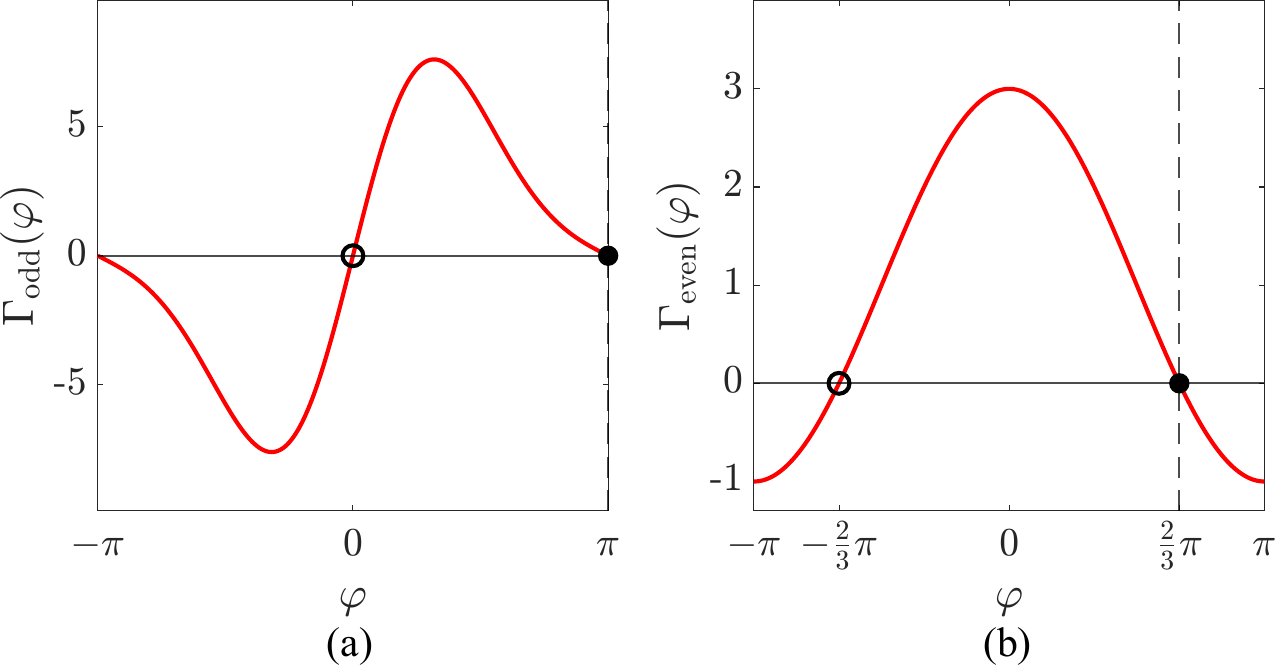}
\caption{
(a)~Odd function $\Gamma_{\mathrm{odd}}$.
(b)~Even function $\Gamma_{\mathrm{even}}$.
In each figure, the black dot shows the stable fixed point and black circle shows the unstable fixed point.
}
\label{fig3}
\end{figure}

We next present the odd function $\Gamma_{\mathrm{odd}}$ and even function $\Gamma_{\mathrm{even}}$ used in this study.
We use an odd function defined by
\begin{align}
\Gamma_{\mathrm{odd}}(\varphi) = 10\sum_{k=1}^{10} k \exp\left( -\frac{k^{2}}{2} \right) \sin(k\varphi),
\end{align}
which is shown in Fig.~\ref{fig3}(a).
This odd function has only one stable fixed point at $\varphi^{*} = \pi$ in $(-\pi,\pi]$, which satisfies the stability condition~\eqref{cond:stability1}.
Also, this odd function satisfies $\Gamma'_{\mathrm{odd}}(\varphi) < 0$ in $\varphi \in [\pi/3,\pi]$
for the stability conditions~\eqref{cond:alpha_sym_stable} and \eqref{cond:beta_sym_stable}.
We define the even function by 
\begin{align}
\Gamma_{\mathrm{even}}(\varphi) = 2\cos(\varphi) + 1,
\end{align}
which is shown in Fig.~\ref{fig3}(b).
This even function has only one stable fixed point at $\varphi^{*} = 2\pi/3$ in $(-\pi,\pi]$, which satisfies the stability condition~\eqref{cond:stability2}.
Also, this even function satisfies $\Gamma'_{\mathrm{even}}(\varphi) < 0$ in $\varphi \in [\pi/3,\pi]$
for the stability conditions~\eqref{cond:alpha_sym_stable} and \eqref{cond:beta_sym_stable}.
We note that, if we consider the transitions between other gaits as given in Appendix A, we need to design other even functions depending on the target fixed point.

\subsection{Gait generation and gait transitions}

We here explain the method for determining the PCF $\Gamma_{1,2}$ and $b_{1,2}$ for the gait transitions of the wave, tetrapod, and tripod gaits.

\subsubsection{Wave gait}

Since the fixed point is $(\alpha^{*},\beta^{*}) = (\pi,\pi/3)$,
we let $\Gamma_{1}$ be the odd function and $\Gamma_{2}$ be the even function:
\begin{align}
\label{cond:wave}
\begin{aligned}
\Gamma_{1}(\alpha) &= \Gamma_{\mathrm{odd}}(\alpha), \\
\Gamma_{2}(\beta) &= -\Gamma_{\mathrm{even}}(\beta + \pi),
\end{aligned}
\end{align}
where $\Gamma_{2}(\beta)$ has a stable fixed point at $\beta^{*} = \pi/3$.
The associated parameters $b_{1,2}$ are determined by
\begin{align}
\begin{aligned}
b_{1} &= 1, \\
b_{2} &= -1
\end{aligned}
\end{align}
from the conditions \eqref{cond:b1} and \eqref{cond:b2}.

\subsubsection{Tetrapod gait}
Since the fixed point is $(\alpha^{*},\beta^{*}) = (2\pi/3,2\pi/3)$ in this case,
we choose both $\Gamma_{1,2}$ as the even functions:
\begin{align}
\label{cond:tetrapod}
\begin{aligned}
\Gamma_{1}(\alpha) &= \Gamma_{\mathrm{even}}(\alpha), \\
\Gamma_{2}(\beta) &= \Gamma_{\mathrm{even}}(\beta).
\end{aligned}
\end{align}
The associated parameters $b_{1,2}$ are determined by
\begin{align}
\begin{aligned}
b_{1} &= -1, \\
b_{2} &= -1
\end{aligned}
\end{align}
from the conditions \eqref{cond:b1} and \eqref{cond:b2}.

\subsubsection{Tripod gait}

Since the fixed point is $(\alpha^{*},\beta^{*}) = (\pi,\pi)$,
we choose both $\Gamma_{1,2}$ as the odd functions:
\begin{align}
\label{cond:tripod}
\begin{aligned}
\Gamma_{1}(\alpha) &= \Gamma_{\mathrm{odd}}(\alpha), \\
\Gamma_{2}(\beta) &= \Gamma_{\mathrm{odd}}(\beta).
\end{aligned}
\end{align}
The associated parameters $b_{1,2}$ are determined by
\begin{align}
\begin{aligned}
b_{1} &= 1, \\
b_{2} &= 1
\end{aligned}
\end{align}
from the conditions \eqref{cond:b1} and \eqref{cond:b2}.


\section{Results of Gait Transitions}

\label{sec:results}

In this section, we show the results of gait transitions in our CPG network by numerical simulations.


\subsection{FitzHugh-Nagumo CPG model}

In our CPG network, various types of limit cycle oscillators can be used as the CPG unit.
We employ the FitzHugh-Nagumo (FHN) oscillator~\cite{Fitzhugh1961impulses,Nagumo1962Active} as the CPG unit, which is a simple mathematical model of a spiking neuron.
The FHN oscillator is described by a two-dimensional vector $\BX = [ x \; y ]^\top$, which obeys the vector field
\begin{align}
\BF(\BX) = d
\begin{bmatrix}
x - a x^{3} - y \\
(x + b) c
\end{bmatrix}
,
\end{align}
where we assume $(a, b, c, d) = (1/3, 0.25, 0.15, 40)$.
The natural period and frequency of this FHN oscillator without scaling ($s = 1$) are $T = 0.548$ and $\omega = 11.5$, respectively.

\begin{figure}
\centering
\includegraphics[width=0.48\textwidth]{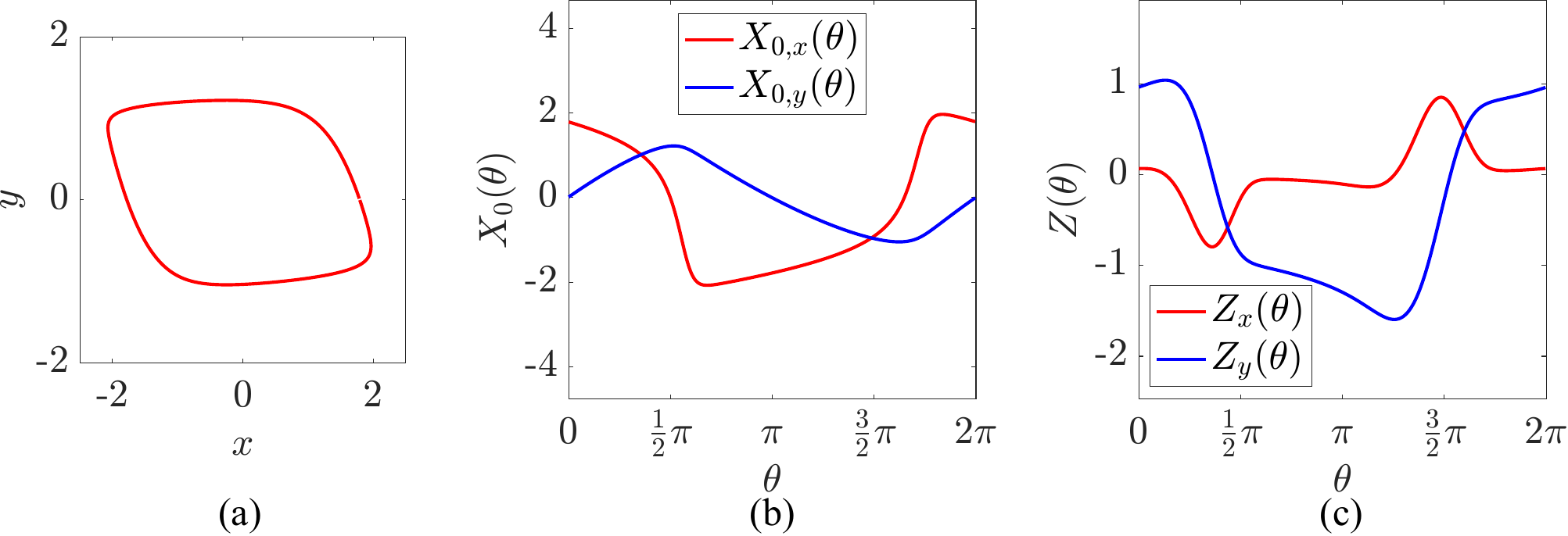}
\caption{
Limit cycle and PSF of the FHN oscillator.
(a)~Limit cycle in the $xy$ plane.
(b)~Limit-cycle solution $\BX_{0} = [X_{0,x}\; X_{0,y}]^{\top}$ vs. phase $\theta$.
(c)~PSF $\BZ = [Z_{x}\; Z_{y}]^{\top}$ vs. phase $\theta$.
}
\label{fig4}
\end{figure}

We show the limit-cycle trajectory in Fig.~\ref{fig4}(a),
each component of the limit-cycle solution in Fig.~\ref{fig4}(b),
and the PSF calculated by the adjoint equation in Fig.~\ref{fig4}(c).
We choose the first component as the output of the CPG unit, i.e., $x_{\mathrm{out}} = x$.
The threshold values $\sigma$ for the wave, tetrapod, and tripod gaits are chosen as $1.6485$, $0.9437$, and $-0.7402$, respectively.

\subsection{Gait transitions}

\subsubsection{Wave-tetrapod-tripod}

We first simulate the gait transitions from wave to tetrapod and tetrapod to tripod in order, where the locomotion speed increases.
The initial gait is the wave gait and it is switched to the tetrapod gait at $t = 12T_{\mathrm{sw}}$, at which the PCFs $\Gamma_{1,2}$ and the binary parameters $b_{1,2}$ are changed.
Then, the tetrapod gait is switched to the tripod gait at $t = 36T_{\mathrm{sw}}$, at which $\Gamma_{1,2}$ and $b_{1,2}$ are changed.
The oscillation timescale $s$ is adjusted according to the rule~\eqref{eq:speed}.
The PCFs $\Gamma_{1,2}$ and $b_{1,2}$ are changed as described in Sec.~\ref{sec:methods}~C.

\begin{figure*}
\centering
\includegraphics[width=\textwidth]{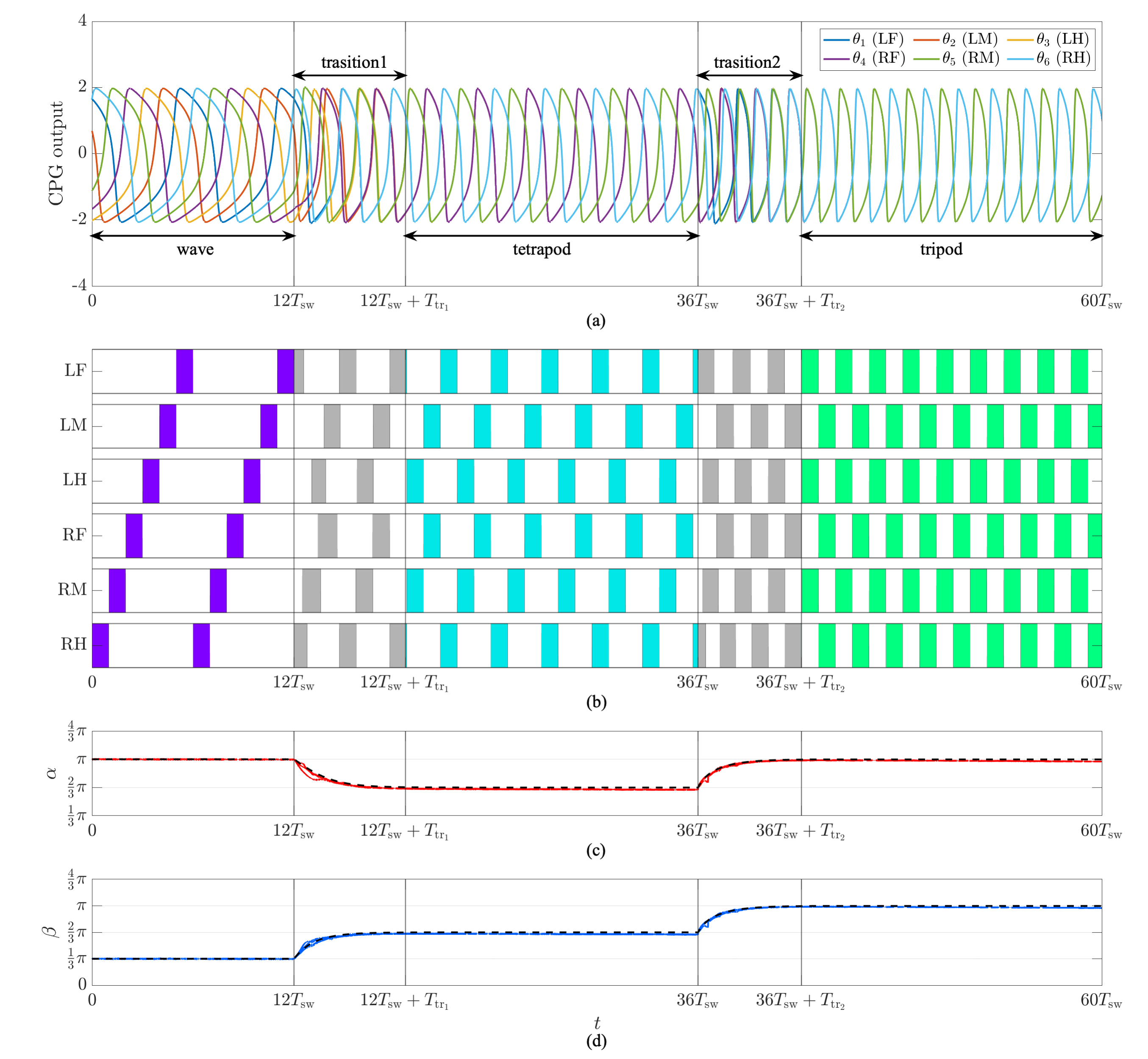}
\caption{
Results for wave-tetrapod-tripod gait transitions.
(a)~Output signals of the CPG network.
The six signals correspond to the individual legs as shown in the legend.
(b)~Gait diagrams over time determined by the rule~\eqref{cond:swing_stance}.
The white area shows the stance phase and the colored bars show the swing phase.
The purple bars show the wave gait, the cyan bars show the tetrapod gait, the green bars show the tripod gait, and the gray bars show the gaits during transitions, respectively.
(c)~Dynamics of the phase difference $\alpha$.
The black dashed curve shows the numerically simulated dynamics of Eq.~\eqref{eq:Gamma_alpha} and 
the red curves show the dynamics of the phase differences $\theta_{k+3} - \theta_{k}\; (k = 1,2,3)$ measured from the oscillator states 
obtained by direct numerical simulations of Eq.~\eqref{eq:model}.
(d)~Dynamics of the phase difference $\beta$.
The black dashed curve shows the numerically simulated dynamics of Eq.~\eqref{eq:Gamma_beta} and 
the blue curves show the dynamics of the phase differences $\theta_{2} - \theta_{1}$, $\theta_{3} - \theta_{2}$, $\theta_{5} - \theta_{4}$, and $\theta_{6} - \theta_{5}$ measured from the directly-simulated oscillator states obeying Eq.~\eqref{eq:model}.
}
\label{fig5}
\end{figure*}

We performed numerical simulations of the dynamics for the phase differences, Eqs.~\eqref{eq:Gamma_alpha}, \eqref{eq:Gamma_beta}, and the original CPG network dynamics, Eq.~\eqref{eq:model}, for the gait transitions
by using the fourth-order Runge-Kutta method with the time interval $\Delta_{t} = 1.0 \times 10^{3}$, where the coupling parameters are chosen as $c_{1} = 4$, $c_{2} = 8$, and $\varepsilon = 0.1$.
We show the outputs of the CPG units in Fig.~\ref{fig5}(a).
We can observe that the amplitudes of the outputs are maintained during the gait transitions,
which does not hinder the swing-stance determination by the threshold value.

The gait diagrams are shown in Fig.~\ref{fig5}(b).
The purple bars show the wave gait, where the legs (RH), (RM), (RF), (LH), (LM), and (LF) are moved in order with the phase shift $\pi/3$. 
The gray bars between the purple and cyan bars show the irregular gait during the transition from the wave gait to the tetrapod gait.
The cyan bars show the tetrapod gait, where the three pairs of legs (LH, RM), (LM, RF), and (LF, RH) are moved in order with the phase shift $2\pi/3$.
The transition time from the wave to tetrapod gait is about $T_{\mathrm{tr}_{1}} = 6.62 \times T_{\mathrm{sw}}$.
The gray bars between the cyan and green bars show the irregular gait during the transition from the tetrapod to tripod gait.
Finally, the green bars show the tripod gait, where the two triplets of legs (RF, LM, RH) and (LF, RM, LH) are moved alternatively with the phase shift $\pi$.
The transition time from the tetrapod gait to the tripod gait is about $T_{\mathrm{tr}_{2}} = 6.13 \times T_{\mathrm{sw}}$.

We show the dynamics of the phase differences $\alpha$ and $\beta$ in Figs.~\ref{fig5}(c) and (d), respectively.
The dynamics of $\alpha$ obtained from Eq.~\eqref{eq:Gamma_alpha} and 
the phase difference $\theta_{k+3} - \theta_{k}\; (k = 1,2,3)$ evaluated from the oscillator states by simulating Eq.~\eqref{eq:model} 
are slightly different during the gait transition from $\alpha = \pi$ to $\alpha = 2\pi/3$, 
but both of them eventually converge to the same target value.
There is also a small difference between $\beta$ obtained from Eq.~\eqref{eq:Gamma_beta} and 
the phase differences $\theta_{2} - \theta_{1}$, $\theta_{3} - \theta_{2}$, $\theta_{5} - \theta_{4}$, and $\theta_{6} - \theta_{5}$
obtained from the oscillator states obeying Eq.~\eqref{eq:model}.
This is because the reduced phase dynamics and original oscillator dynamics subjected to weak perturbations are slightly different,
causing a small breakdown of the symmetries, but the phases eventually converge to the target phase and recover the symmetries because of the stability.
Thus, the present CPG network can achieve the wave-tetrapod-tripod gait transitions 
by using the approximate phase-difference dynamics as designed. 

\subsubsection{Tripod-tetrapod-wave}

Next, we simulate the tripod to tetrapod and tetrapod to wave transitions in order, where the locomotion speed decreases.
The gait is initially the tripod gait and it is switched to the tetrapod gait at $t = 12T_{\mathrm{sw}}$.
Then, the tetrapod gait is switched to the wave gait at $t = 36T_{\mathrm{sw}}$.
The oscillation timescale $s$ is adjusted according to the rule~\eqref{eq:speed}.
The gait transition rule is the same as that in the previous case.

\begin{figure*}
\centering
\includegraphics[width=\textwidth]{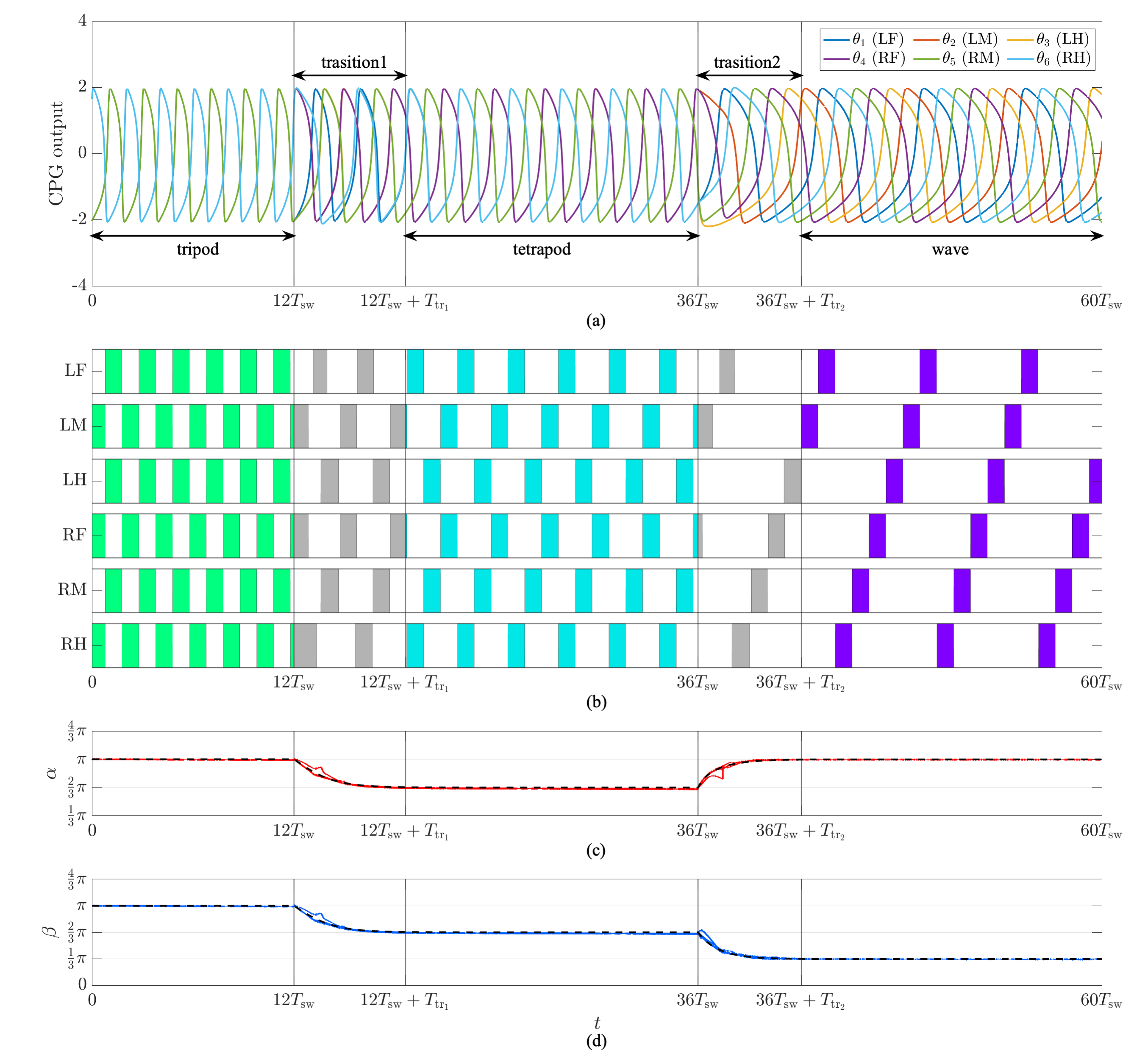}
\caption{
Results for tripod-tetrapod-wave gait transitions.
(a)~Output signals of the CPG network.
The six signals correspond to the individual legs as shown in the legend.
(b)~Gait diagrams over time determined by the rule~\eqref{cond:swing_stance}.
The white area shows the stance phase and the colored bars show the swing phase.
The green bars show the tripod gait, the cyan bars show the tetrapod gait, the purple bars show the wave gait, and the gray bars show the gaits during transitions.
(c)~Dynamics of the phase difference $\alpha$.
The black dashed curve shows the numerically simulated dynamics of Eq.~\eqref{eq:Gamma_alpha} and 
the red curves show the dynamics of the phase differences $\theta_{k+3} - \theta_{k}\; (k = 1,2,3)$ measured from the oscillator states 
obtained by direct numerical simulations of Eq.~\eqref{eq:model}, respectively.
(d)~Dynamics of the phase difference $\beta$.
The black dashed curve shows the numerically simulated dynamics of Eq.~\eqref{eq:Gamma_beta} and 
the blue curves show the dynamics of the phase differences $\theta_{2} - \theta_{1}$, $\theta_{3} - \theta_{2}$, $\theta_{5} - \theta_{4}$, and $\theta_{6} - \theta_{5}$ measured from the oscillator states obtained by direct numerical simulations of Eq.~\eqref{eq:model}.
}
\label{fig6}
\end{figure*}

We performed numerical simulations of the gait transitions by using the same procedure as described before.
We show the outputs of the CPG network in Fig.~\ref{fig6}(a), where the amplitude of the outputs are maintained during the gait transitions.

We show the gait diagrams in Fig.~\ref{fig6}(b).
The green bars show the tripod gait, the cyan bars show the tetrapod gait, and the gray bars between the green and cyan bars show the irregular gait during the transition from the tripod to tetrapod gait, respectively.
The transition time from the tripod gait to the tetrapod gait is about $T_{\mathrm{tr}_{1}} = 6.62 \times T_{\mathrm{sw}}$.
From this and the previous results, we can observe that the tetrapod gait can be achieved from both the wave and tripod gaits.

The purple bars show the wave gait, and the gray bars between the cyan and purple bars show the irregular gait during the transition from the tetrapod to wave gait, respectively.
The transition time from the tetrapod to the wave gait is about $T_{\mathrm{tr}_{2}} = 6.13 \times T_{\mathrm{sw}}$.

We show the evolution of the phase differences $\alpha$ and $\beta$ in Figs.~\ref{fig6}(c) and (d), respectively.
The dynamics of $\alpha$ obtained from Eq.~\eqref{eq:Gamma_alpha} and 
the phase difference $\theta_{k+3} - \theta_{k}\; (k = 1,2,3)$ evaluated from the oscillator states by simulating Eq.~\eqref{eq:model} 
are slightly different during the gait transitions, but they eventually converge to the same target value.
There are also slight differences between the dynamics of $\beta$ obtained from Eq.~\eqref{eq:Gamma_beta} and 
the phase differences $\theta_{2} - \theta_{1}$, $\theta_{3} - \theta_{2}$, $\theta_{5} - \theta_{4}$, and $\theta_{6} - \theta_{5}$ evaluated from the oscillator states by simulating Eq.~\eqref{eq:model},
but all of them eventually converge to the same target value.
This is due to the slight difference between the reduced phase dynamics and original oscillator dynamics subjected to weak perturbations,
which causes a small breakdown of the symmetries, but the phases eventually go to the target phase and recover the symmetries from the stability.
Thus, the present CPG network can also achieve the tripod-tetrapod-wave gait transitions.

\section{Conclusions}

\label{sec:conclusions}

In this study, we presented a simple CPG network that can control the gait transitions in hexapod between the different types of gaits whose symmetries are described in Sec.~\ref{sec:model}~B.
In our CPG network, the dynamics of the six CPG oscillators can be reduced to the dynamics of two independent phase differences
between the contralateral legs on the left and right sides and between the adjacent ipsilateral legs on both sides.
By appropriately assuming the functional form of the mutual coupling functions,
the desired synchronization dynamics can be achieved irrespective of the detailed properties of the oscillators.
Also, since the coupling between the oscillators is assumed weak, the output signals of the CPGs maintain approximately constant amplitudes and do not hinder the determination of the swing-stance phase. 
We demonstrated that our CPG network can achieve the wave-tetrapod-tripod gait transitions (speed increasing) and tripod-tetrapod-wave gait transitions (speed decreasing) smoothly by numerical simulations.
The present CPG network is also applicable to other gait patterns in Appendix~\ref{sec:gait_variations}, which possess the symmetries described in Sec.~\ref{sec:model}~B.

The advantage of the present CPG framework is the ability to provide a general CPG network solely using two independent phase-difference dynamics. 
Therefore, the performance of the gait transitions can easily be optimized, compared to most existing studies that rely on specific models of the CPG oscillators.
At the same time, since the present framework allows us to use various types of limit-cycle oscillators as the CPG unit, 
we can employ not only existing models of oscillators but also artificially designed oscillators~\cite{Isjpeert2013dynamical,Ajalooeian2013general,Grzelczyk2016prototype,Namura2023designing,Namura2023design} as the CPG unit.
In our future work, we will focus on optimizing the coupling functions within the present CPG network to achieve faster gait transitions.


\acknowledgments
We acknowledge financial support from JSPS KAKENHI (Nos. JP22K11919 and JP22H00516) and JST CREST (No. JPMJCR1913). 


\appendix

\section{Gait variations}
\label{sec:gait_variations}

In the main text, we considered three typical types of symmetric gaits, i.e., the wave, tetrapod, and tripod gaits.
However, there exist other gait patterns that are also called the ``wave'' or ``tetrapod'' gait in the literature.
We here list them up to the best of our knowledge. 
Also, we describe other gait patterns than the wave, tetrapod, and tripod gaits.
We show that all gaits in this Appendix possess the symmetries described in Sec.~\ref{sec:model}~B,
which can be controlled by using the two independent phase differences of our CPG network.


\subsection{Wave gaits}

There are many gait patterns called the ``wave'' gait, such as
(i) (LF), (RF), (LM), (RM), (LH), (RH)~\cite{Yu2016gait},
(ii) (LF), (LM), (LH), (RF), (RM), (RH)~\cite{Bal2021neural} (also called the metachronal tripod in Ref.~\cite{Inagaki2006wave} or metachronal$^{+}$ in Ref.~\cite{Golubitsky1998modular}),
and (iii) (LF), (RM), (LH), (RF), (LM), (RH)~\cite{Minati2018versatile} (also called the reversed rolling tripod in Ref.~\cite{Inagaki2006wave} or rolling tripod$^{-}$ in Ref.~\cite{Golubitsky1998modular}), 
where (iv) (LF), (RH), (LM), (RF), (LH), (RM) is considered as a similar gait to (iii),
which is called the rolling tripod in Ref.~\cite{Inagaki2006wave} or rolling tripod$^{+}$ in Ref.~\cite{Golubitsky1998modular}.
We show the phase shifts when the phase of LF is $\theta_{1} = 0$ and fixed points $(\alpha^{*},\beta^{*})$ corresponding to these gaits in Table~\ref{tab:wave}.

\begin{table*}
\centering
\caption{Other patterns of wave gaits.}
\label{tab:wave}
\tabcolsep = 5pt
\renewcommand{\arraystretch}{1.5}
\begin{tabular}{cccc}
\hline \hline
& gait patterns & $(\theta_{1},\theta_{2},\theta_{3},\theta_{4},\theta_{5},\theta_{6})$ & $(\alpha^{*},\beta^{*})$ \\
\hline
(i): & (LF), (RF), (LM), (RM), (LH), (RH) & $\left( 0, \frac{4}{3}\pi, \frac{2}{3}\pi, \frac{5}{3}\pi, \pi, \frac{1}{3}\pi \right)$ & $\left( \frac{5}{3}\pi, \frac{4}{3}\pi \right)$ \\
(ii): & (LF), (LM), (LH), (RF), (RM), (RH) & $\left( 0, \frac{5}{3}\pi, \frac{4}{3}\pi, \pi, \frac{2}{3}\pi, \frac{1}{3}\pi \right)$ & $\left( \pi, \frac{5}{3}\pi \right)$ \\
(iii): & (LF), (RM), (LH), (RF), (LM), (RH) & $\left( 0, \frac{2}{3}\pi, \frac{4}{3}\pi, \pi, \frac{5}{3}\pi, \frac{1}{3}\pi \right)$ & $\left( \pi, \frac{2}{3}\pi \right)$ \\
(iv): & (LF), (RH), (LM), (RF), (LH), (RM) & $\left( 0, \frac{4}{3}\pi, \frac{2}{3}\pi, \pi, \frac{1}{3}\pi, \frac{5}{3}\pi \right)$ & $\left( \pi, \frac{4}{3}\pi \right)$ \\
\hline \hline
\end{tabular}
\end{table*}


\subsection{Tetrapod gaits}

In the main text, we considered the tetrapod gait given by (LH, RM), (LM, RF), (LF, RH), but there can be other types of tetrapod gait patterns~\cite{Aminzare2018gait}, i.e.,
(i) (LF, RM), (LH, RF), (LM, RH) called the forward left tetrapod,
(ii) (LF, RH), (LM,RF), (LH, RM) called the backward right tetrapod, 
and
(iii) (LF, RM), (LM, RH), (LH, RF) called the backward left tetrapod.
Moreover, there exists another tetrapod pattern (iv), in which one or two legs of (LF,RH), (RM), (LH,RF), (LM)~\cite{Chen2012smooth,Minati2018versatile,Bal2021neural} are moved simultaneously at equal phase interval $\pi/2$,
which corresponds to an intermediate speed gait between the wave and tetrapod gait considered in the main text~\cite{Wilson1966insect}.
We show the phase shifts when the phase of LF is $\theta_{1} = 0$ and the fixed points $(\alpha^{*},\beta^{*})$ corresponding to these gaits in Table~\ref{tab:tetrapod}.

\begin{table*}
\centering
\caption{Other patterns of tetrapod gaits.}
\label{tab:tetrapod}
\tabcolsep = 5pt
\renewcommand{\arraystretch}{1.5}
\begin{tabular}{cccc}
\hline \hline
& gait patterns & $(\theta_{1},\theta_{2},\theta_{3},\theta_{4},\theta_{5},\theta_{6})$ & $(\alpha^{*},\beta^{*})$ \\
\hline
(i): & (LF, RM), (LH, RF), (LM, RH) & $\left( 0, \frac{2}{3}\pi, \frac{4}{3}\pi, \frac{4}{3}\pi, 0, \frac{2}{3}\pi \right)$ & $\left( \frac{4}{3}\pi, \frac{2}{3}\pi \right)$ \\
(ii): & (LF, RH), (LM, RF), (LH, RM) & $\left( 0, \frac{4}{3}\pi, \frac{2}{3}\pi, \frac{4}{3}\pi, \frac{2}{3}\pi, 0 \right)$ & $\left( \frac{4}{3}\pi, \frac{4}{3}\pi \right)$ \\
(iii): & (LF, RM), (LM, RH), (LH, RF) & $\left( 0, \frac{4}{3}\pi, \frac{2}{3}\pi, \frac{2}{3}\pi, 0, \frac{4}{3}\pi \right)$ & $\left( \frac{2}{3}\pi, \frac{4}{3}\pi \right)$ \\
(iv): & (LF, RH), (RM), (LH, RF), (LM) & $\left( 0, \frac{3}{2}\pi, \pi, \pi, \frac{1}{2}\pi, 0 \right)$ & $\left( \pi, \frac{3}{2}\pi \right)$ \\
\hline \hline
\end{tabular}
\end{table*}


\subsection{Other gait patterns}

Furthermore, we can consider other gait patterns than the wave, tetrapod, and tripod gaits,
e.g., walking with (i) all legs moving at the same time (called pronk),
(ii) with two triplets of legs (LF, LM, LH) and (RF, RM, RH) moved alternatively (called pace),
(iii) with two sets of legs (LF, LH, RF, RH) and (LM, RM) moved alternatively (called lurch),
(iv) with three pairs of legs (LF, RF), (LM, RM) and (LH, RH) moved in order, where some pause exists between the pairs (LH, RH) and (LF, RF), which is called inchworm$^{+}$, 
(v) with three pairs of legs (LF, RF), (LH, RH), and (LM, RM) moved in order, where some pause exists between the pairs (LH, RH) and (LF, RF), which is called inchworm$^{-}$,
(vi) with three pairs of legs (LF, RF), (LM, RM), and (LH, RH) moved at equal phase intervals (called caterpillar$^{+}$),
and (vii) with three pairs of legs (LF, RF), (LH, RH), and (LM, RM) moved at equal phase intervals (called caterpillar$^{-}$)~\cite{Golubitsky1998modular}.
We show the phase shifts when the phase of LF is $\theta_{1} = 0$ and the fixed points $(\alpha^{*},\beta^{*})$ corresponding to these gaits in Table~\ref{tab:other_gaits}.

\begin{table*}
\centering
\caption{Gait patterns other than the wave, tetrapod, and tripod gaits.}
\label{tab:other_gaits}
\tabcolsep = 5pt
\renewcommand{\arraystretch}{1.5}
\begin{tabular}{cccc}
\hline \hline
& gait patterns & $(\theta_{1},\theta_{2},\theta_{3},\theta_{4},\theta_{5},\theta_{6})$ & $(\alpha^{*},\beta^{*})$ \\
\hline
(i): & (LF,LM,LH,RF,RM,RH) & $\left( 0, 0, 0, 0, 0, 0 \right)$ & $\left( 0, 0 \right)$ \\
(ii): & (LF,LM,LH), (RF,RM,RH) & $\left( 0, 0, 0, \pi, \pi, \pi \right)$ & $\left( \pi, 0 \right)$ \\
(iii): & (LF,LH,RF,RH), (LM,RM) & $\left( 0, \pi, 0, 0, \pi, 0 \right)$ & $\left( 0, \pi \right)$ \\
(iv): & (LF,RF), (LM,RM), (LH,RH) & $\left( 0, \frac{5}{3}\pi, \frac{4}{3}\pi, 0, \frac{5}{3}\pi, \frac{4}{3}\pi \right)$ & $\left( 0, \frac{5}{3}\pi \right)$ \\
(v): & (LF,RF), (LH,RH), (LM,RM) & $\left( 0, \frac{1}{3}\pi, \frac{2}{3}\pi, 0, \frac{1}{3}\pi, \frac{2}{3}\pi \right)$ & $\left( 0, \frac{1}{3}\pi \right)$ \\
(vi): & (LF,RF), (LM,RM), (LH,RH) & $\left( 0, \frac{4}{3}\pi, \frac{2}{3}\pi, 0, \frac{4}{3}\pi, \frac{2}{3}\pi \right)$ & $\left( 0, \frac{4}{3}\pi \right)$ \\
(vii): & (LF,RF), (LH,RH), (LM,RM) & $\left( 0, \frac{2}{3}\pi, \frac{4}{3}\pi, 0, \frac{2}{3}\pi, \frac{4}{3}\pi \right)$ & $\left( 0, \frac{2}{3}\pi \right)$ \\
\hline \hline
\end{tabular}
\end{table*}


\bibliographystyle{unsrt}
\bibliography{references}

\end{document}